\documentclass[prb,twocolumn,amsmath,amssymb]{revtex4}
\usepackage{graphicx}
\usepackage{dcolumn}
\usepackage{bm}
\def\Xint#1{\mathchoice
   {\XXint\displaystyle\textstyle{#1}}%
   {\XXint\textstyle\scriptstyle{#1}}%
   {\XXint\scriptstyle\scriptscriptstyle{#1}}%
   {\XXint\scriptscriptstyle\scriptscriptstyle{#1}}%
   \!\int}
\def\XXint#1#2#3{{\setbox0=\hbox{$#1{#2#3}{\int}$}
     \vcenter{\hbox{$#2#3$}}\kern-.5\wd0}}
 \def\dashint{\Xint-}

\voffset=.7truecm \hoffset=-.4truecm \newcommand{\Tc}{$T_{c}$}

\begin{document}
\title{Superconductivity-Induced Transfer of In-Plane Spectral Weight in Bi$_{2}$Sr$_{2}$CaCu$_{2}$O$_{8}$:
Resolving a Controversy}
\author{A. B. Kuzmenko, H. J. A. Molegraaf, F. Carbone, and D. van der Marel}
\affiliation{D\'epartement de Physique de la Mati\`ere
Condens\'ee, Universit\'ee de Gen\`eve, CH-1211 Gen\`eve 4,
Switzerland}

\date{\today}

\begin{abstract} We present a detailed analysis of the
superconductivity-induced redistribution of optical spectral
weight in Bi$_{2}$Sr$_{2}$CaCu$_{2}$O$_{8}$ near optimal doping.
It confirms the previous conclusion by Molegraaf {\it et al.}
(Science {\bf 295}, 2239 (2002)), that the integrated
low-frequency spectral weight shows an extra increase below \Tc.
Since the region, where the change of the integrated spectral
weight is not compensated, extends well above 2.5 eV, this
increase is caused by the transfer of spectral weight from
interband to intraband region and only partially by the narrowing
of the intraband peak. We show that the opposite assertion by
Boris {\it et al.} (Science {\bf 304}, 708 (2004)), regarding this
compound, is unlikely the consequence of any obvious discrepancies
between the actual experimental data. \end{abstract}

\maketitle
\section{Introduction}
The problem of the superconductivity (SC) induced spectral weight
(SW) transfer in the high-\Tc\ cuprates is in the focus of
numerous experimental
\cite{MolegraafScience02,SantanderEPL03,BorisScience04,SchuetzmannPRB97,MolerScience98,
TsvetkovNature98,KirtleyPRL99,GaifullinPRL99,BasovScience99,BorisPRL02,KuzmenkoPRL03,
MarelKluwer04,DeutscherCM05} and theoretical
\cite{HirschPC92,AndersonScience95,LeggettScience96,ChakravartyEPJB98,
ChakravartyPRL99,MunzarPRB01,MunzarPRB03,NormanPRB02,MarelKluwer03,KnigavkoPRB04,
AbanovPRB04,BenfattoEPJB04,AndersonJPCM04,HirschPRB04,StanescuPRB04,MaierPRL04,FaridPM04}
studies. Molegraaf {\it et al.} \cite{MolegraafScience02} reported
the SC induced increase of the intraband spectral weight in
optimally doped (\Tc\ = 88 K) and underdoped (\Tc\ = 66 K) single
crystals of Bi$_{2}$Sr$_{2}$CaCu$_{2}$O$_{8}$ (Bi2212) based on
combined ellipsometry and reflectivity measurements.
Santander-Syro {\it et al.} \cite{SantanderEPL03} independently
arrived at the same conclusions by analyzing reflectivity spectra
of high-quality Bi2212 films. However, these results were recently
questioned by Boris {\it et al.} \cite{BorisScience04}, who
reported fully ellipsometric measurements of optimally doped
YBa$_{2}$Cu$_{3}$O$_{6.9}$ (Y123, \Tc\ = 92.7 K) and slightly
underdoped Bi2212 (\Tc\ = 86 K) and concluded that in these two
materials there is 'a sizable superconductivity-induced decrease
of the total intraband spectral weight'. It is thus necessary to
clarify this experimental controversy.

The following reasons can potentially cause the opposite
conclusions by different teams, namely: (i) non-identical
compounds and doping levels were used, (ii) essentially different
experimental results were obtained, or (iii) the analysis and
interpretation of similar experimental data seriously deviate,
most likely, in the way to determine the SC-caused change of the
low-frequency SW:
\begin{equation}\label{W}
W(\Omega_{c},T)=\rho_{s}(T)+\int_{0^+}^{\Omega_{c}}\sigma_{1}(\omega,T)d\omega,
\end{equation}
where $\rho_{s}(T)$ is the spectral weight of the
condensate, $\sigma_{1}(\omega,T)$ is the real part of the optical
conductivity and the cutoff energy $\Omega_{c}$ represents the
scale of scattering of the intraband (Drude) excitations.

Because of reason (i), we restrict our present discussion to the
case of Bi2212 near optimal doping, which was studied by all
mentioned teams
\cite{BorisScience04,MolegraafScience02,SantanderEPL03}. We should
keep in mind that, even with this restriction, the \Tc's and the
stoichiometries of the samples are still not identical.

Regarding point (ii) we found that the experimental graphs,
presented in Ref. \onlinecite{BorisScience04}, do not
significantly deviate from our earlier observations on optimally
doped Bi2212 \cite{MolegraafScience02}. In particular, we show
that the data on Bi2212, presented in Ref.
\onlinecite{BorisScience04}, given the error bars, do not
challenge the previous conclusion about the SC-induced increase of
$W(\Omega_{c},T)$ $(\Omega_{c}$ was 1.25 eV in
Ref.\onlinecite{MolegraafScience02}).

This leaves possibility (iii) as the only remaining option: The
contradicting conclusions of Ref. \onlinecite{BorisScience04} and
those of Refs. \onlinecite{MolegraafScience02,SantanderEPL03}
originate from a different analysis of the mutually
non-contradicting experimental data. In the present paper we
scrutinize the arguments of Ref. \onlinecite{BorisScience04} and
we point out an unjustified use of the Kramers-Kronig (KK)
relations in interpreting their data, which appears to be
responsible, at least in part, for the conclusions opposite to
Refs. \onlinecite{MolegraafScience02} and
\onlinecite{SantanderEPL03}.

The limited space of Ref.\onlinecite{MolegraafScience02} did not
allow explaining in depth the full analysis done. In this paper we
present a more detailed and rigorous analysis of the experimental
data published in Ref.\onlinecite{MolegraafScience02} and arrive
at additional arguments supporting the original conclusions. In
order to numerically decouple the superconductivity-induced
changes of the optical properties from the temperature
dependencies, already present in the normal state (such as a
gradual narrowing of the Drude peak), we apply the
slope-difference analysis, developed in Ref.
\onlinecite{KuzmenkoPRL03}, which is closely related to the
well-known temperature-modulation technique
\cite{Cardona69,HolcombPRL94}. It appears that any realistic
model, which satisfactorily fits the total set of experimental
data (reflectivity below 0.75 eV and ellipsometrically obtained
real and imaginary parts of the dielectric function at higher
energies), gives an {\it increase} of $W(\Omega_{c}, T)$ below
\Tc. Moreover, all attempts to do the same fitting with an
artificial constraint, that the SC-induced change of
$W(\Omega_{c}, T)$ is negative (as is claimed in
Ref.\onlinecite{BorisScience04}), or even zero, failed to fit the
data, despite using flexible multi-oscillator
models\cite{KuzmenkoCM05}. Importantly, the ellipsometrically
measured real and imaginary parts of the dielectric function,
$\epsilon_{1}(\omega)$ and $\epsilon_{2}(\omega)$, {\it at high
frequencies} provide the most stringent limits on the possible
change of the {\it low-frequency} spectral weight $W(\Omega_{c},
T)$, due to the KK relations.

Finally we discuss whether the SC-increase of $W(\Omega_{c}, T)$
is caused by the extra narrowing of the Drude peak below \Tc\ or
by removal of the SW from the interband region. Indeed we find an
extra narrowing of the Drude peak, in agreement with
Ref.\onlinecite{BorisScience04}. However, this narrowing is too
small to explain the increase of the low-frequency spectral
weight, which suggests that there is a sizeable spectral weight
transfer from the range of the interband transitions.
\section{Experimental consistency of the reported results}
\label{Consistency}
\begin{figure}[thb]
   \includegraphics[width=8cm,clip=true]{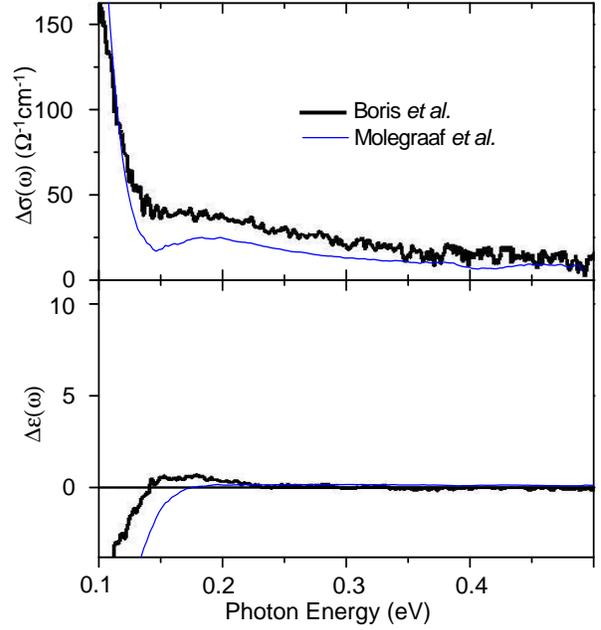}
   \caption{Differential dielectric function
   $\Delta\epsilon_{1}(\omega)=\epsilon_{1}(\omega,100 K)-\epsilon_{1}(\omega,20
   K)$ and optical conductivity $\Delta\sigma_{1}(\omega)=\sigma_{1}(\omega,100 K)-\sigma_{1}(\omega,20
   K)$ of Bi2212 near optimal doping, measured by ellipsometry in
   Ref.\onlinecite{BorisScience04} (black curve)
   and derived by the KK transformation of reflectivity data from Ref. \onlinecite{MolegraafScience02} (blue curve).
   The latter KK-transformed data are shown for comparison only; our analysis
   is based on the directly measured reflectivity in the infrared and ellipsometry above 0.75 eV.
   }
   \label{Fig1}
\end{figure}

First of all, we need to check if there are any qualitative
discrepancies between the results published in
Ref.\onlinecite{BorisScience04} and our data, which can
immediately lead to the opposite conclusions. In Fig.\ref{Fig1} we
reproduce the difference curves
$\Delta\epsilon_{1}(\omega)=\epsilon_{1}(\omega,100
K)-\epsilon_{1}(\omega,20 K)$ and optical conductivity
$\Delta\sigma_{1}(\omega)=\sigma_{1}(\omega,100
K)-\sigma_{1}(\omega,20 K)$ for Bi2212 close to optimal doping
from ellipsometric measurements of Ref.\onlinecite{BorisScience04}
(Fig. S5) together with the same curves, obtained by the KK
transformation of the reflectivity spectra of Ref.
\onlinecite{MolegraafScience02}, re-plotted in the same fashion.
Although for the purpose of comparison of spectra in this range we
have to use the KK-transformed quantities, the main analysis which
we present in sections IV and V of this paper is based on the {\it
directly measured} optical quantities. We also point out right
away, that these difference curves do not eliminate the
normal-state temperature trends, unrelated to the SC phase
transition. In sections IV and V we present what we believe to be
the correct analysis, which takes into account the normal-state
temperature dependence of optical properties.

One can state that there is a qualitative agreement between the
$\Delta\epsilon_{1}(\omega)$ and $\Delta\sigma_{1}(\omega)$ curves
of the two groups, especially considering that these difference
curves are on a 10-fold amplified scale as compared to the spectra
itself (Fig. 1 of Ref. \onlinecite{MolegraafScience02}, Fig. S4 of
Ref. \onlinecite{BorisScience04}) and a possible difference of the
sample compositions.

Clearly, no model-independent argument, without specific
calculations, can give opposite signs of the change of the SW,
when applied to these two data sets. Therefore, the different
conclusions of different teams are unlikely the consequence of any
obvious discrepancies between the actual experimental data.

An important remark is that the scale and the data scatter on
Fig.\ref{Fig1} do not allow to distinguish the change of
$\epsilon_{1}(\omega)$ from zero for frequencies between 0.2 and
0.5 eV. As a result, the authors of
Ref.\onlinecite{BorisScience04} base their argument on
$\Delta\epsilon_{1}(\omega)$ being zero in this range. In
contrast, later we show that it is small, but not zero, including
the range above 0.5 eV. In fact, this observation will be
extremely important to determine the sign of the SW change.
\section{On the role of the Kramers-Kronig relations}\label{KK}
Since the optical conductivity $\sigma_{1}(\omega)$ is not
directly measured down to zero frequency, the calculation of
$W(\Omega_{c},T)$ requires, in principle, the extrapolation of
data, using certain data modelling. It turns out, however, that
the use of the Kramers-Kronig relation between
$\epsilon_{1}(\omega)$ and $\sigma_{1}(\omega)$
\begin{equation} \epsilon_{1}(\omega) - 1 =
8\dashint_{0}^{\infty}\frac{\sigma_{1}(x)}{x^{2}-\omega^{2}}dx
\end{equation}
\noindent helps to avoid unnecessary model assumptions and
drastically decrease the resulting error bars.

The authors of Ref.\onlinecite{BorisScience04} propose a {\it
model-independent} argument in order to show that there is an
overall decrease of the intraband spectral weight below \Tc. The
argument is based on two of their experimental
observations\cite{signconvention}: (i) $\Delta\sigma_{1}(\omega) >
0$ between $\omega_{1}$ = 0.15 eV and $\omega_{2}$ = 1.5 eV, (ii)
$\Delta\epsilon_{1}(\omega)=0$ in the same energy region. We
reproduce partially their statement here: {\it "The SW loss
between 0.15 eV and 1.5 eV then needs to be balanced by a
corresponding SW gain below 0.15 eV and above 1.5 eV. In other
words, there is necessarily a corresponding SW gain in the
interband energy range above 1.5 eV caused by a decrease of the
total intraband SW."}

We believe that this statement is incorrect. The justification
given by the authors of Ref.\onlinecite{BorisScience04} considers
a weaker set of conditions: $\Delta\sigma_{1}(\omega) > 0$ for
$\omega_{1} < \omega < \omega_{2}$, and
$\Delta\epsilon_{1}(\omega_{0}) = 0$ at one particular frequency
$\omega_{0}$ between $\omega_{1}$ and $\omega_{2}$. We rewrite
Equation (S3) of Ref. \onlinecite{BorisScience04} in the form:
\begin{eqnarray}
-\int_{0}^{\omega_{1}}\frac{\Delta\sigma_{1}(x)}{\left|x^2-
\omega_{0}^2\right|}
+\dashint_{\omega_{1}}^{\omega_{2}}\frac{\Delta\sigma_{1}(x)}{x^2-
\omega_{0}^2}\nonumber\\
+\int_{\omega_{2}}^{\infty}\frac{\Delta\sigma_{1}(x)}{\left|x^2-
\omega_{0}^2\right|} = -A + B + C = 0.
\end{eqnarray}
\noindent Now we note, that even though $\Delta\sigma_{1}$ in the
integral $B$ is positive, the value of $B$ can have either sign.
It means that there is formally no limitation on the sign of
$C-A$, nor on the signs of $A$ and $C$ separately. Appealing to
the general $f$-sum rule
$\int_{0}^{\infty}\Delta\sigma_{1}(x)dx=0$ does not remove the
uncertainty, because the change of the low-frequency spectral
weight can be, in principle, compensated at arbitrarily high
frequencies, which give an arbitrarily small contribution to the
KK integral.

The stronger condition that $\Delta\epsilon_{1}(\omega)$ is
exactly zero for $\omega_{1} < \omega < \omega_{2}$ is physically
pathological and it appears to be more difficult to address
mathematically. It was not explained in
Ref.\onlinecite{BorisScience04} how this condition leads to the
cited statement. Furthermore, we observe that
$\Delta\epsilon_{1}(\omega)$ is actually {\it not zero} in this
range. We thus disagree with the model-independent argument in
favor of the SC-induced decrease of the charge carrier SW, as is
claimed in Ref.\onlinecite{BorisScience04}.

On the other hand, we absolutely agree with
Ref.\onlinecite{BorisScience04}, that the KK relations between
$\epsilon_{1}(\omega)$ and $\sigma_{1}(\omega)$ must stay as an
important ingredient of the data analysis \cite{Text1}. However,
in our opinion trustworthy conclusions can only be drawn from a
thorough numerical treatment of the full set of available optical
data. Importantly, due to the KK relations, the behavior of
$\epsilon_{1}(\omega,T)$ and $\epsilon_{2}(\omega,T)$ at high
frequencies appears to be the most sensitive indicator of the
spectral weight transfer. We present such an analysis in the
following sections.
\section{Superconductivity-related spectral changes}\label{Kinks}
As described in Ref.\onlinecite{MolegraafScience02}, the
normal-incidence reflectivity $R(\omega)$ was measured between 200
and 6000 cm$^{-1}$ (25 meV - 0.75 eV) and real and imaginary parts
of dielectric function $\epsilon_{1}(\omega)$ and
$\epsilon_{2}(\omega)$, were obtained by spectroscopic
ellipsometry from 6000 to 36000 cm$^{-1}$ (0.75 - 4.5 eV).

\begin{figure}[thb]
   \includegraphics[width=8cm,clip=true]{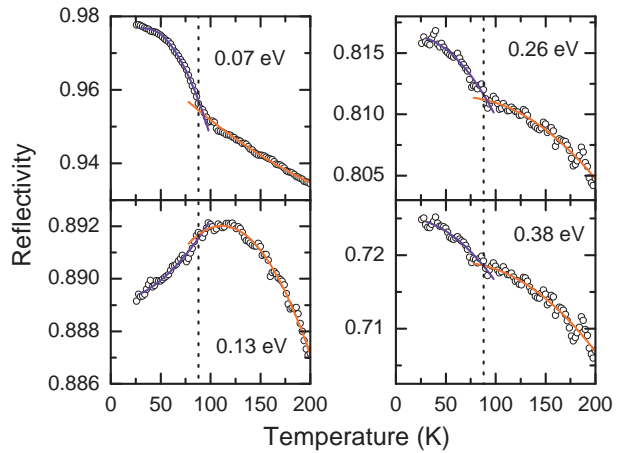}
   \caption{Temperature dependent reflectivity of Bi2212 close to optimal
   doping (\Tc = 88 K) for selected frequencies in the infrared range.
   The blue (red) curves are polynomial fits to the temperature dependence
   below (above) \Tc, used to produce the kink (slope difference) values.}
   \label{Fig2}
\end{figure}

\begin{figure}[thb]
   \includegraphics[width=8cm,clip=true]{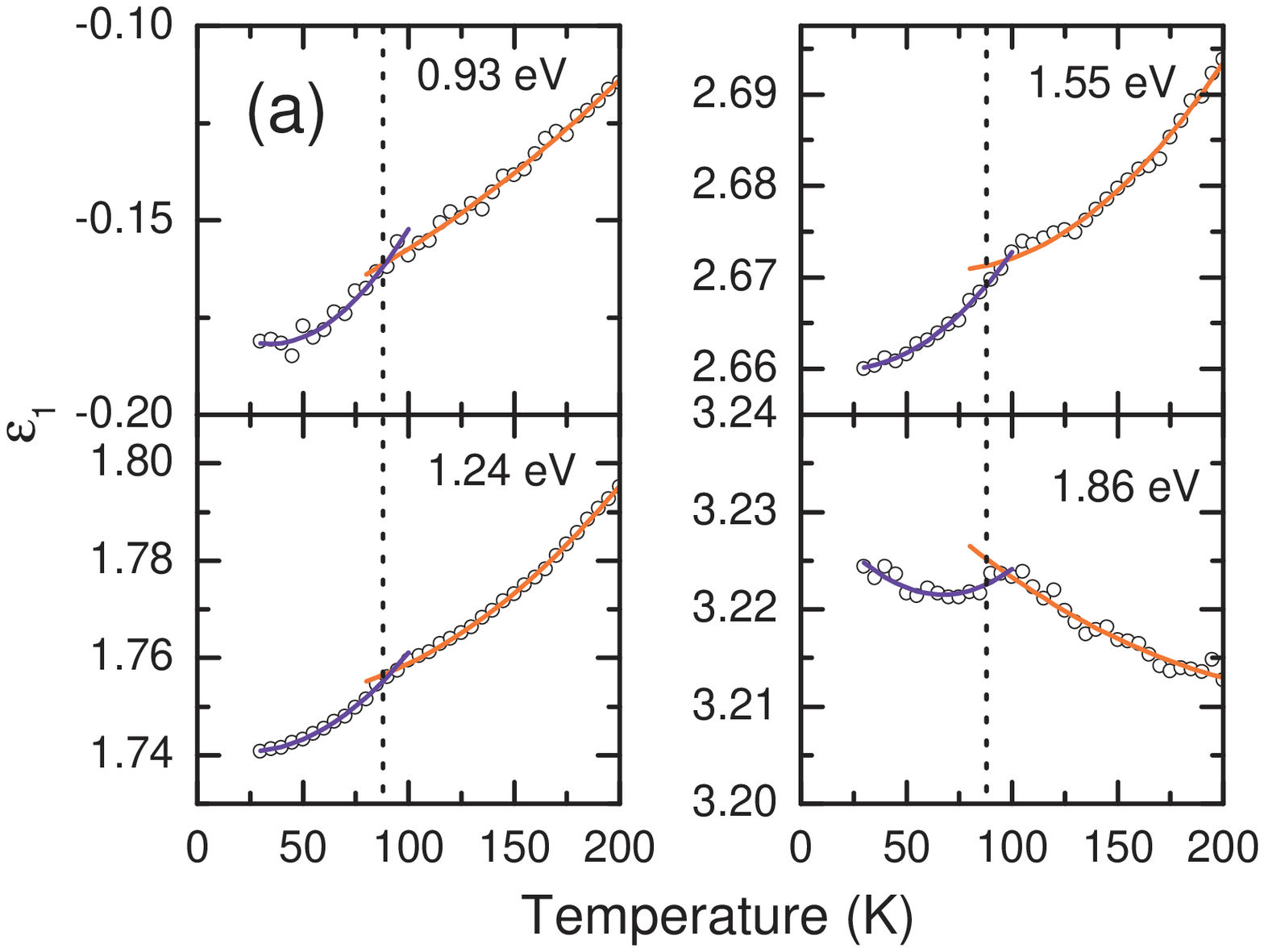}
   \includegraphics[width=8cm,clip=true]{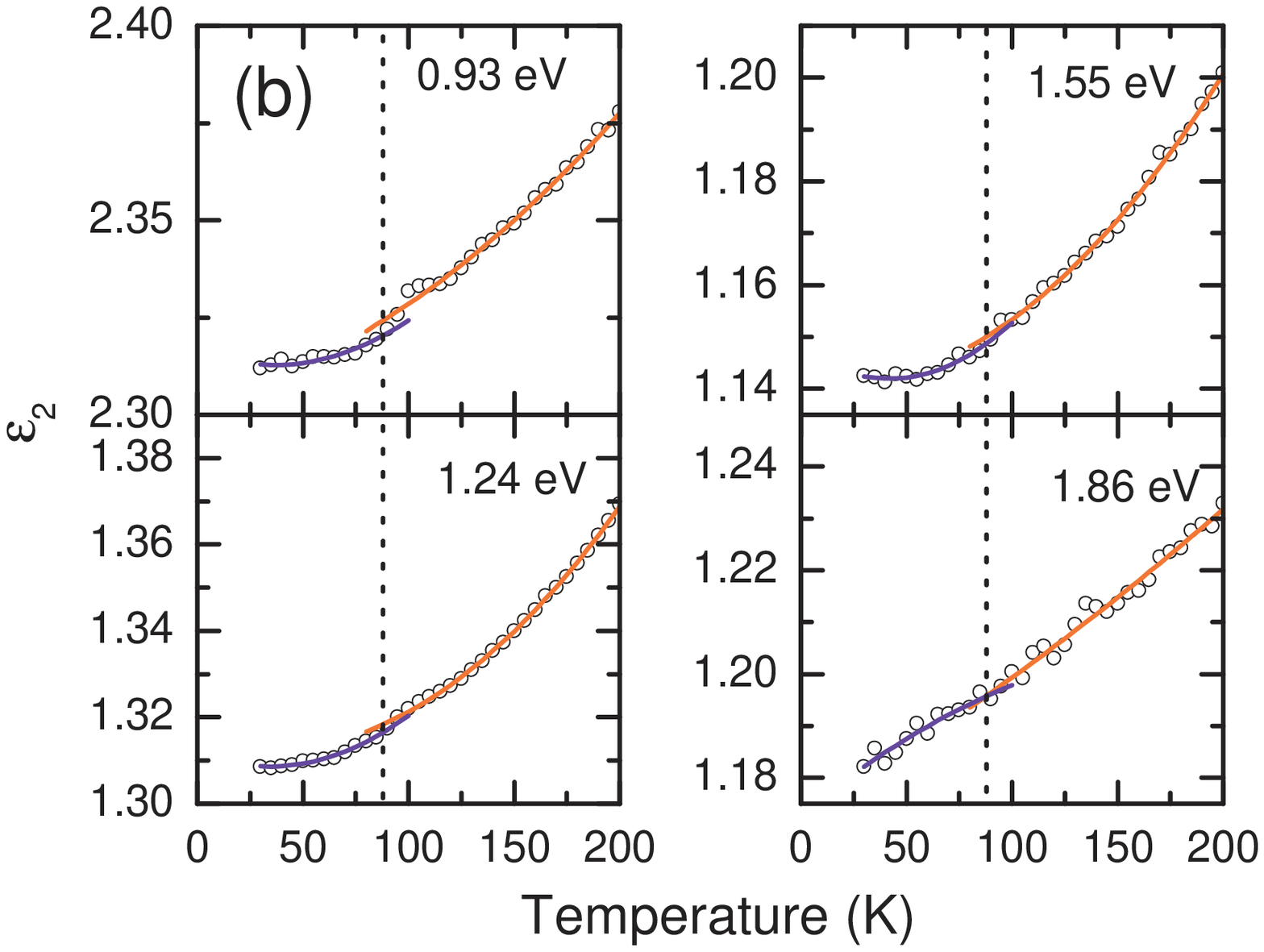}
   \caption{Temperature dependent ellipsometrically measured
   $\epsilon_{1}(T)$ (a) and $\epsilon_{2}(T)$ (b) of Bi2212 close to optimal
   doping (\Tc = 88 K) for selected frequencies. The blue (red)
   curves are polynomial fits to the temperature dependence below (above) \Tc, used to
   produce the kink (slope change) values. Not all temperature
   datapoints are shown.}
   \label{Fig3}
\end{figure}

In Figs. \ref{Fig2}, \ref{Fig3} we display the temperature
dependence of reflectivity, and $\epsilon_{1,2}(\omega,T)$
respectively, for selected photon energies. The complex dielectric
constant changes as a function of temperature in the entire range
from 0 to 300 K. For frequencies larger than 0.25 eV the variation
as a function of temperature is essentially proportional to
$T^{2}$. The same temperature dependence has been observed in
other cuprate superconductors, for example\cite{ortolani05}
La$_{2-x}$Sr$_{x}$CuO$_{4}$, and has been explained quantitatively
using the Hubbard model\cite{ToschiCM05}. For frequencies below
0.1 eV the optical conductivity has a very large
intraband-contribution, with a linearly increasing dissipation,
which causes a non-monotonous temperature dependence of
$\epsilon_{1}$ and $\epsilon_{2}$. In particular, it was found
\cite{MarelNature03} that the $1/(T\sigma_{1}(\omega))$ is equal
to a constant plus a term proportional to $T^{-2}$. The gradual
decrease of the high-frequency conductivity with cooling down is
also expected due to the reduction of the electron-phonon
scattering\cite{KarakozovSSC02}.

The onset of the superconductivity is marked by clear kinks (slope
changes) at \Tc\ of the measured optical quantities
(Fig.\ref{Fig2}, \ref{Fig2}). An exciting feature of the high-\Tc\
cuprates is that the kinks are seen not only in the region of the
SC gap, but also at much higher photon energies (at least up to
2.5 eV in Ref. \onlinecite{MolegraafScience02}). It means that the
formation of the SC long-range order causes a redistribution of
the spectral weight across a very large spectral range; a fact,
which several groups agree upon
\cite{BorisScience04,MolegraafScience02,SantanderEPL03,RuebhausenPRB01,HolcombPRL94}.

An essential aspect of the data analysis is the way to separate
the superconductivity-induced changes of the optical constants
from the temperature-dependent trends, observed above \Tc. A
similar problem is faced in the specific-heat experiments
\cite{JunodPC99,LoramJPCS01}, where the superconductivity-related
structures are superimposed on a strong temperature dependent
background. Let us introduce a slope-difference operator
$\Delta_{s}$, which measures the slope change (kink) at \Tc
\cite{KuzmenkoPRL03}:
\begin{equation}\label{kinkdef} \Delta_{s} f(\omega) \equiv
\left.\frac{\partial f(\omega,T)}{\partial T} \right|
_{T_{c}+\delta}-\left.\frac{\partial f(\omega,T)}{\partial T}
\right| _{T_{c}-\delta}, \end{equation} \noindent where $f$ stands
for any optical quantity. It properly quantifies the effect of the
SC transition, since the normal state trends are cancelled
out\cite{footnotekink}. Since $\Delta_{s}$ is linear, the
slope-difference KK relation is also valid:
\begin{equation}\label{kkslope} \Delta_{s}\epsilon_{1}(\omega)=
8\dashint_{0}^{\infty}\frac{\Delta_{s}\sigma_{1}(x)}{x^{2}-\omega^{2}}dx.
\end{equation}

In order to detect and measure a kink as a function of temperature
as well as to establish whether or not the kink is seen at \Tc\,
the spectra must be measured using a fine temperature resolution.
A resolution of 2 K or better was used in the entire range from 10
to 300 K, enables us to perform a reliable slope-difference
analysis.

\begin{figure*}[thb]
   \includegraphics[width=8cm,clip=true]{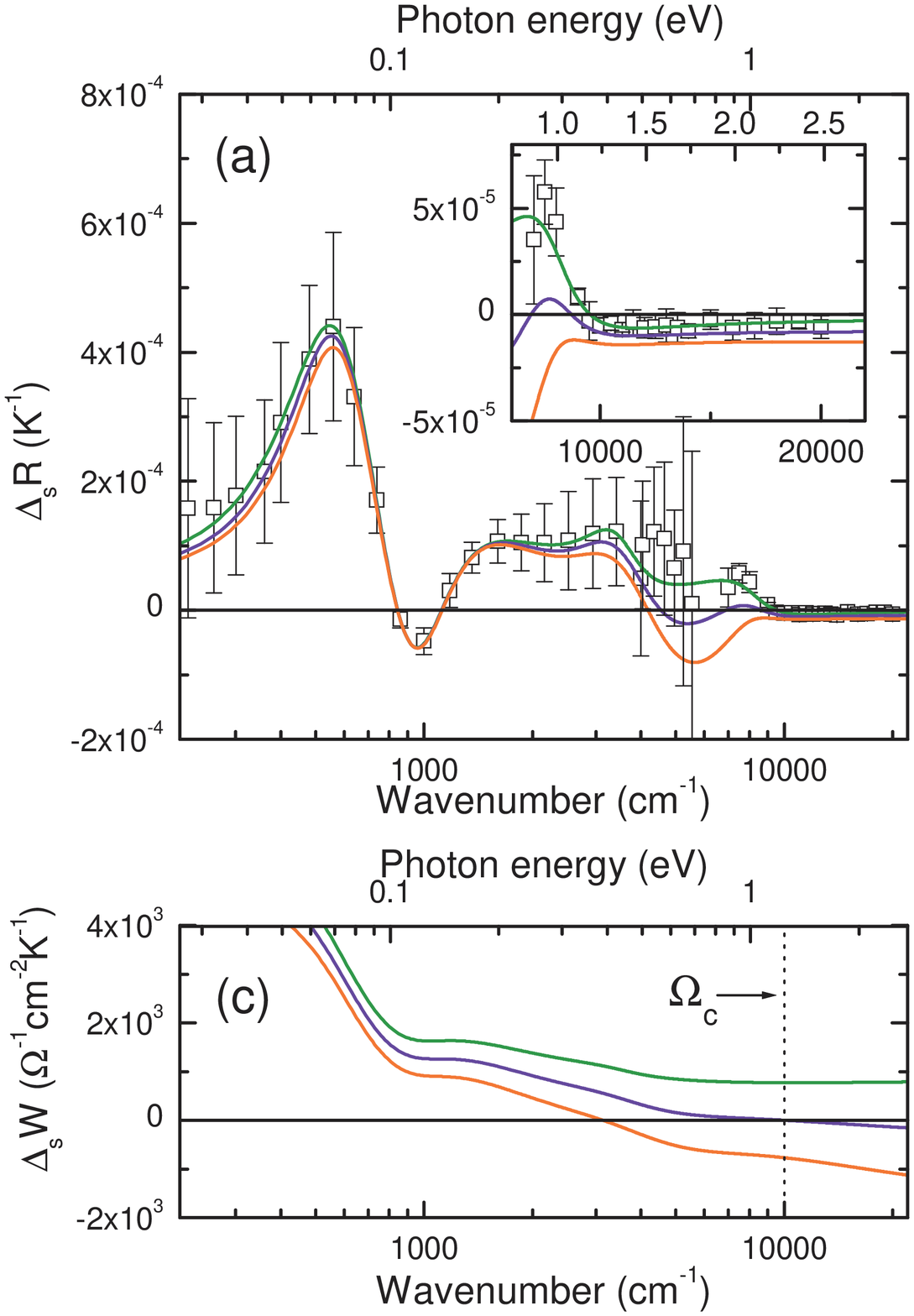}
   \includegraphics[width=8cm,clip=true]{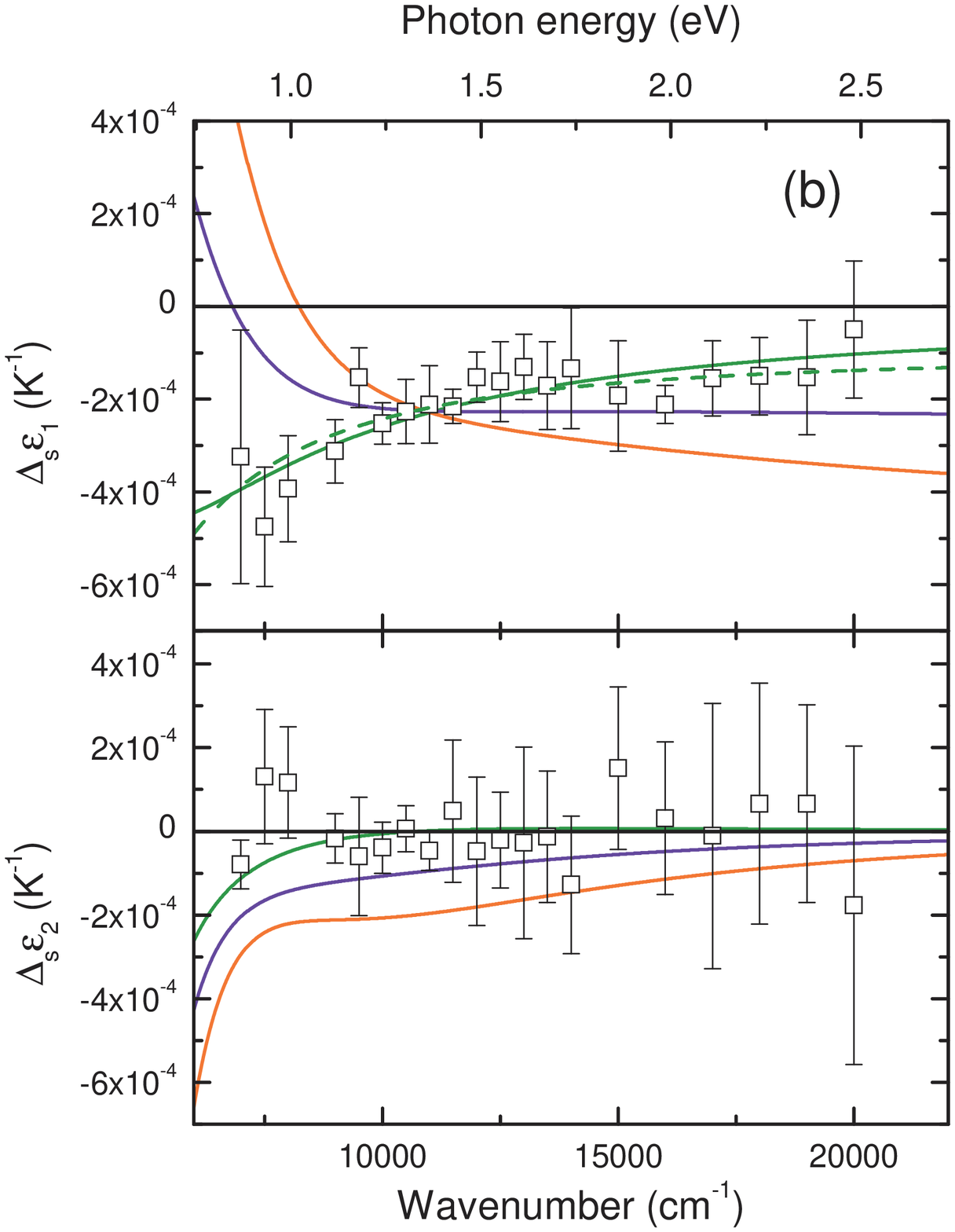}
   \caption{The spectral dependence of slope-difference $\Delta_{s}R(\omega)$ (a) and
   $\Delta_{s}\epsilon_{1,2}(\omega)$ and $\Delta_{s}\epsilon_{2}(\omega)$ (b).
   Datapoints show the kink value, derived from the temperature-dependent
   curves (Figs.\ref{Fig2} and \ref{Fig3}) as described in
   Appendix A1. Only few datapoints of $\Delta_{s}R(\omega)$ out of all used in the analysis are shown.
   Solid curves show different fits, as described in the text.
   The green curve indicates the best fit of $\Delta_{s}R(\omega)$, $\Delta_{s}\epsilon_{1}(\omega)$ and
   $\Delta_{s}\epsilon_{2}(\omega)$ simultaneously, giving $\Delta_{s}W(\Omega_{c})$
   $\approx$ +770 $\Omega^{-1}$cm$^{-2}$K$^{-1}$ (the SC-induced {\it increase} of
   $W(\Omega_{c})$). The blue and red curves are the best fits of the same data with an artificial
   constraint $\Delta_{s}W(\Omega_{c})$ = 0 and -770 $\Omega^{-1}$cm$^{-2}$K$^{-1}$ respectively.
   One can see that both constraints are incompatible with the data.
   The green dashed line is the best fit of $\Delta_{s}\epsilon_{1}(\omega)$
   by the formula $-A\omega^{-2}+B$.  The integrated spectral weight $\Delta_{s}W(\omega)$ (c) for all three
   fits is shown by the same colors. The inset of panel (a) shows
   $\Delta_{s}R(\omega)$ calculated from the ellipsometrically
   measured dielectric function from 0.75 to 2.75 eV on an
   expanded vertical scale.
   }
   \label{Fig4}
\end{figure*}

The datapoints in Figs. \ref{Fig4}a and \ref{Fig4}b show
$\Delta_{s}R(\omega)$ and  $\Delta_{s}\epsilon_{1}(\omega)$ and
$\Delta_{s}\epsilon_{2}(\omega)$, obtained from the directly
measured temperature-dependent curves shown in Fig. \ref{Fig2} and
\ref{Fig3}. The details of the corresponding numerical procedure
and the determination of the error bars are described in Appendix
A1.

One can see, that $\Delta_{s}\epsilon_{1}(\omega)$ is {\it
negative} and its absolute value is strongly decreasing as a
function of frequency. In the same region,
$\Delta_{s}\epsilon_{2}(\omega)$ is almost zero (within our error
bars). Intuitively, it already suggests that the low-frequency
integrated spectral weight $W(\Omega_{c},T)$ is likely to increase
in the SC state. Indeed, in the simplest scenario, when the extra
SW is added at zero frequency, one has
$\Delta_{s}\epsilon_{1}(\omega)=-8\Delta_{s}W(0+)\omega^{-2}$.
This formula becomes approximate, if changes also take place at
finite frequencies. However, the approximation is good, if the
most significant changes $\Delta_{s}\sigma_{1}$ occur only below
$\omega_{1} \ll \omega$, and above $\omega_{2} \gg \omega$. This
follows from the exact expansion\cite{BozovicPRB90} of Eq.
\ref{kkslope}, valid for $\omega_{1}<\omega<\omega_{2}$:
\begin{eqnarray} \Delta_{s}\epsilon_{1}(\omega) =
-\sum_{j=0}^{\infty}\frac{A_{j}}{\omega^{2+2j}} +
\Delta_{s}\tilde{\epsilon}_1(\omega)
+\sum_{j=0}^{\infty}B_{j}\omega^{2j}, \label{swap}
\end{eqnarray}
\noindent
where
\begin{eqnarray}
A_{j}&\equiv&8\int_{0}^{\omega_{1}}x^{2j}\Delta_{s}\sigma_{1}(x)dx \nonumber\\
B_{j}&\equiv&8\int_{\omega_{2}}^{\infty}\frac{\Delta_{s}\sigma_{1}(x)}{x^{2+2j}}dx\nonumber\\
\Delta_{s}\tilde{\epsilon}_{1}(\omega)
&\equiv&8\dashint_{\omega_{1}}^{\omega_{2}}\frac{\Delta_{s}\sigma_{1}(x)}{x^{2}-\omega^{2}}dx.
\nonumber \end{eqnarray} \noindent
For $\omega_{1}$ = 0.8 eV and $\omega_{2}$ = 2.5 eV we can
calculate $\Delta_{s}\tilde{\epsilon}_{1}(\omega)$ directly from
the measured $\Delta_{s}\sigma_{1}(\omega)$. It turns out, that
$\left|\Delta_{s}\tilde{\epsilon}_{1}(\omega)\right|$ $<$
10$^{-4}$ K$^{-1}$, while the average value is indistinguishable
from zero. In this situation we can neglect
$\Delta_{s}\tilde{\epsilon}_{1}(\omega)$, compared to the
contributions from the low and high frequencies. Thus, to leading
order
\begin{eqnarray}\label{swapprox}
\Delta_{s}\epsilon_{1}(\omega)&\approx&-A_{0}\omega^{-2}+B_{0}\nonumber\\
&=&-8\Delta_{s}W(\omega_{1})\omega^{-2}+\Delta_{s}\epsilon_{\infty},
\end{eqnarray} \noindent
where
$\epsilon_{\infty}=8\int_{\omega_{2}}^{\infty}\sigma_{1}(\omega)\omega^{-2}d\omega$
is the integrated oscillator strength of all optical excitations
above $\omega_{2}$.

The best fit of $\Delta_{s}\epsilon_{1}(\omega)$ in the spectral
region (0.8 - 2.5 eV) with Eq. \ref{swapprox} is shown by the
green dashed line in Fig.\ref{Fig4}b. It gives
$\Delta_{s}W(\Omega_{c})$ $\approx$ +360
$\Omega^{-1}$cm$^{-2}$K$^{-1}$ and
$\Delta_{s}\epsilon_{\infty}$$\approx$ -10$^{-4}$ K$^{-1}$.
Although these absolute values are very approximate, the {\it
signs} of both parameters indicate that the spectral weight is
taken from the region above 2.5 eV and added to the region below
0.8 eV.

\section{Slope-difference spectral analysis} \label{DiffAnalysis}
Now we present the full data analysis without using the
approximation (\ref{swapprox}). The technique we present here is a
modification of the temperature-modulation
spectroscopy\cite{Cardona69,HolcombPRL94,DolgovPC94}. Because of
the KK relation (\ref{kkslope}), we can model the slope-difference
dielectric function with the following dispersion formula:
$\Delta_{s}\epsilon(\omega)$:
\begin{equation}\label{DL}
\Delta_{s}\epsilon(\omega)=\Delta_{s}\epsilon_{\infty}+
\sum_{i=0}^{N}\frac{A_{i}}{\omega_{i}^2-\omega^{2}-i\gamma_{i}\omega}.
\end{equation}
\noindent $\Delta_{s}\epsilon_{\infty}$ is responsible for the
high-frequency electronic excitations, while each Lorentzian term
represents either an addition or a removal of spectral weight,
depending on the sign of $A_{i}$. We emphasize that this model is
just a parametrization. The number of oscillators $N$ has to be
chosen in order to get a good fit of experimental data. The
physical meaning of some oscillators, taken alone, may not be
well-defined. However, the essential feature of the functional
form (\ref{DL}) is that it preserves\cite{Text6} the KK relation
(\ref{kkslope}).

We include the infrared $\Delta_{s}R(\omega)$ to the fitting
procedure by making use of the following
relation\cite{KuzmenkoPRL03,DolgovPC94}:
\begin{eqnarray}
\Delta_{s}R(\omega)= \frac{\partial R}{\partial
\epsilon_{1}}(\omega,T_{c}) \Delta_{s}\epsilon_{1}(\omega)+
\frac{\partial R}{\partial \epsilon_{2}}(\omega,T_{c})
\Delta_{s}\epsilon_{2}(\omega)\nonumber\\  = 2R(\omega,T_{c})
\mbox{ Re}\left[
\frac{\Delta_{s}\epsilon(\omega)}{\sqrt{\epsilon(\omega,T_{c})}\left(\epsilon(\omega,T_{c})-1\right)}\right].
\end{eqnarray}
\noindent The method which we use to determine the 'sensitivity
functions' $(\partial R/\partial \epsilon_{1,2})(\omega,T_{c})$
from the experimental data is described in Appendix A2.

The green solid line in Fig.\ref{Fig4}a and \ref{Fig4}b denotes
the best fit\cite{REFFIT} of $\Delta_{s}\epsilon_{1}(\omega)$ and
$\Delta_{s}\epsilon_{2}(\omega)$ and, {\it simultaneously},
$\Delta_{s}R(\omega)$. One can see that all essential spectral
details are well reproduced. The corresponding parameter values
are collected in Table \ref{TableDeps}. The first term in
(\ref{DL}) combines the condensate and the narrow ($\gamma$ $<$
100 cm$^{-1}$) quasi-particle peak, while the remaining
oscillators mimic the redistribution of spectral weight at finite
frequencies.

\begin{table}
\caption{Model parameters of $\Delta_{s}\epsilon(\omega)$ by
formula (\ref{DL}) which correspond to the best fit of the
experimental data. $\Delta_{s}\epsilon_{\infty}$ =
-3.1$\cdot$10$^{-5}$ K$^{-1}$.}
\begin{tabular}{crrr}
  \hline\hline

  $i$ &  \ \ \ \ $\omega_{i}$ (cm$^{-1}$)  & $A_{i}$ (10$^{5}$ cm$^{-2}$K$^{-1}$) & $\gamma_{i}$ (cm$^{-1}$)\\
  \hline
0 & 0    &   2.82 &      0 \\
1 & 0    & -11.25 &   3977 \\
2 & 582  &  -2.40 &    488 \\
3 & 939  &   2.20 &    707 \\
4 & 2078 &   6.51 &   4496 \\
5 & 2082 &   2.60 &   6141 \\
6 & 3543 &  -0.17 &   1685 \\
    \hline\hline
\end{tabular}
\label{TableDeps}
\end{table}

The slope-difference integrated spectral weight for the model
(\ref{DL}):
\begin{equation}
\Delta_{s}W(\omega)=\frac{A_{0}}{8}+\int_{0+}^{\omega}\Delta_{s}\sigma_{1}(x)dx
\end{equation}
\noindent is presented as a green curve in Fig.\ref{Fig4}c. It
gives $\Delta_{s}W(\Omega_{c})$ $\approx$ +770
$\Omega^{-1}$cm$^{-2}$K$^{-1}$, which is about two times larger
than the rough estimate in Section IV.

In order to test how robust this result is, we did two more fits
of the same data, with an extra imposed constraint, that either
(i) $\Delta_{s}W(\Omega_{c})$ = 0, or (ii)
$\Delta_{s}W(\Omega_{c})$ = -770 $\Omega^{-1}$cm$^{-2}$K$^{-1}$.
The resulting 'best-fitting' curves are shown in Fig.\ref{Fig4}a,
\ref{Fig4}b and \ref{Fig4}c in blue and red, respectively. One can
clearly see that the models with $\Delta_{s}W(\Omega_{c})\le 0$
fail to reproduce the experimental spectra, most spectacularly the
high-frequency spectrum of $\Delta_{s}\epsilon_{1}(\omega)$. This
is not surprising, since the imposed constraint changes the sign
of the leading term $\sim\omega^{-2}$ in formula (\ref{swapprox}).

To exclude the possibility that the failure to get a good fit with
the mentioned constraint would be a spurious result caused by the
limited number of Lorentz oscillators used for fitting the data,
we have also used the KK-constrained variational dielectric
model\cite{KuzmenkoCM05}, which is, in simple terms, a collection
of a large number of adjustable oscillators, uniformly distributed
in the whole spectral range, including the region above 2.5 eV.
Thus we are confident that the dispersion model is able to
reproduce all significant spectral features of the true function
$\Delta_{s}\epsilon_{1}(\omega)$. However, these efforts did not
improve the quality of the fit.

From this analysis we conclude that our experimental data {\it
unequivocally} reveal the superconductivity-induced increase of
$W(\Omega_{c},T)$ in optimally doped Bi2212, confirming the
statements given in
Refs.\onlinecite{MolegraafScience02,SantanderEPL03}.

An alternative approach is to fit the total set of spectra at
every temperature\cite{MolegraafScience02,HajoThesis} using a
certain KK-consistent model with temperature-dependent parameters.
There is a variety of possibilities to parameterize the dielectric
function. However, it was shown by one of us\cite{HajoThesis} that
every model which reproduces satisfactorily not only the spectral
features, but also the temperature dependence of the directly
measured $R(\omega)$ and $\epsilon(\omega)$, gives a net increase
of the low-frequency spectral weight below \Tc.
\section{Discussion} \label{Discussion}
\subsection{Absolute change of the spectral weight}
Having found that the integrated spectral weight $W(\Omega_{c},T)$
exhibits an extra {\it increase} below \Tc, we want to evaluate
its {\it absolute} superconductivity-induced change, continued
into the low-temperature region\cite{signconvention}:
\begin{equation} \Delta W(\Omega_{c},T) \equiv W(\Omega_{c},T) -
W_{n}(\Omega_{c},T),
\end{equation} \noindent
where $W_{n}(\Omega_{c},T)$ is the 'correct' extrapolation of the
normal-state curve below \Tc. In principle, $W_{n}(\Omega_{c},T)$
can be measured, when the superconducting order parameter is
suppressed by an extremely high magnetic field (of the order of
hundred Teslas). Unfortunately, such large fields are currently
prohibitive for accurate optical experiments.

However, an order-of-magnitude estimate and, simultaneously, an
upper limit of $\Delta W(\Omega_{c}, T)$ at zero temperature can
be obtained by the formula
\begin{equation}\label{absolute} \Delta W(\Omega_{c}, 0) \sim
T_{c} \Delta_{s}W(\Omega_{c}).
\end{equation} \noindent
This gives $\Delta W(\Omega_{c}, 0K)\sim 7\cdot 10^{4}\
\Omega^{-1}$cm$^{-2}$, which is about 1\% of the total
low-frequency spectral weight $W(\Omega_{c}, 0K)$$\approx$ 7
$\cdot$ 10$^{6}$ $\Omega^{-1}$cm$^{-2}$. Since the temperature
dependence of $\Delta W(\Omega_{c}, T)$ is expected to saturate
somewhat below $T_c$, a more realistic estimate is smaller by
about a factor of 2 to 5, {\it i.e.} between 0.2 and 0.5\% of
$W(\Omega_{c}, 0K)$. These rough margins are suggested by the
temperature dependence of the ab-plane\cite{LeePRL96} and
c-axis\cite{GaifullinPRL99} penetration depths. Although this is a
relatively small fraction, it is nevertheless
significant\cite{MolegraafScience02} in the context of the
theories where the superconducting transition is driven by the
lowering of the kinetic energy
\cite{HirschPC92,AndersonScience95}.
\subsection{The origin of the spectral weight transfer}
The low-frequency integrated spectral weight $W(\Omega_{c},T)$
should not be confused with the intraband spectral weight:
\begin{equation}\label{Wintra}
W_{intra}(T)=\rho_{s}(T)+\int_{0+}^{\infty}\sigma_{1,intra}(\omega,T)d\omega,
\end{equation} \noindent where $\sigma_{1,intra}(\omega,T)$ is the
conductivity due to intraband transitions. A legitimate question
is whether the observed increase of $W(\Omega_{c},T)$ below \Tc\
is due to the transfer of SW from the interband transitions to the
intraband (Drude) conductivity {\it i. e.}, by the increase of
$W_{intra}(T)$, or is simply caused by an extra narrowing of the
Drude peak in the SC state\cite{BorisScience04,KarakozovSSC02},
without changing $W_{intra}(T)$. The distinction between the
intraband and interband spectral weights can be made in
theoretical models, but in experimental spectra their separation
is not unique, because of the unavoidable overlap between these
spectral ranges. Nevertheless, the temperature and spectral
behavior of the optical constants suggests a likely scenario.

\begin{figure}[thb]
   \includegraphics[width=8cm,clip=true]{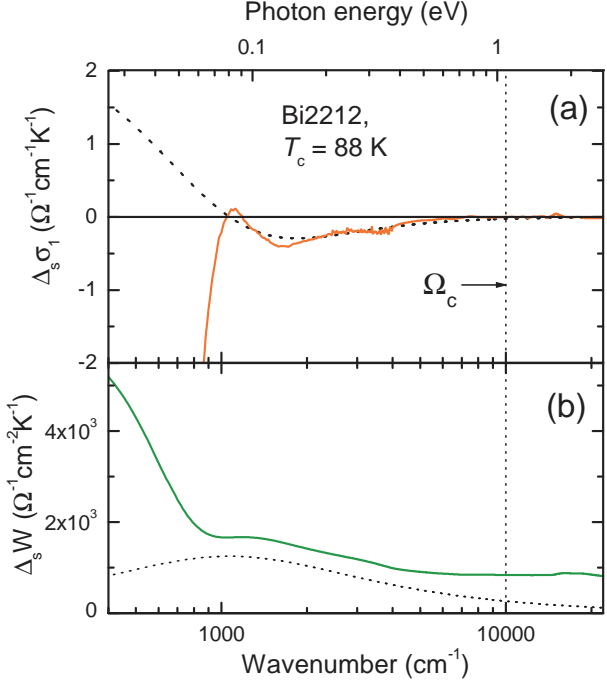}
   \caption{(a) Slope-difference conductivity
   $\Delta_{s}\sigma_{1}(\omega)$ of optimally doped Bi2212. (b)
   The corresponding integrated spectral weight
   $\Delta_{s}W(\omega)$. Dotted lines show a model, which mimics
   the narrowing of the Drude peak.
   }
   \label{Fig5}
\end{figure}

Fig.\ref{Fig5}a shows the slope-difference conductivity
$\Delta_{s}\sigma_{1}(\omega)$, which is obtained by the most
detailed multi-oscillator fit\cite{KuzmenkoCM05} of the data,
shown in Fig.\ref{Fig4}. The decrease of $\sigma_{1}(\omega)$
below \Tc\ in the frequency range from 0.15 to about 0.8 eV is due
to an extra narrowing of the Drude peak, caused by the reduced
charge carrier scattering in the superconducting state. In the
most crude model where the intraband peak is described by the
Drude formula
$\sigma_{1}(\omega)=(4\pi)^{-1}\omega_{p}^2\gamma(\omega^{2}+\gamma^{2})^{-1}$
and the superconductivity-induced narrowing is given by a simple
change of $\gamma$, we get
$\Delta_{s}\sigma_{1}(\omega)=(4\pi)^{-1}\omega_{p}^2(\omega^{2}-\gamma^{2})
(\omega^{2}+\gamma^{2})^{-2}\Delta_{s}\gamma$. This shape (with
$\gamma$ = 0.14 eV) matches the experimental curve above 0.15 eV
quite well (the blue dashed line), even though the real shape of
the conductivity peak is much more complicated. Below 0.12 eV the
drop of the conductivity is caused by the opening of the SC gap.
Thus, a peak of $\Delta_{s}\sigma_{1}(\omega)$ at 0.13 - 0.14 eV,
where the SC-induced change of $\sigma_{1}$ is even positive, is
probably a cooperative effect of the narrowing of the Drude peak
and the suppression of conductivity in the gap region. This peak
corresponds to a dip in the spectrum $\Delta_{s}R(\omega)$
(Fig.\ref{Fig4}a).

Fig.\ref{Fig5}b depicts $\Delta_{s}W(\omega)$, which is obtained
by the integration of $\Delta_{s}\sigma_{1}(\omega)$ of
Fig.\ref{Fig5}a. There is almost no superconductivity-induced
change of $\sigma_{1}(\omega)$ above 0.8 eV up to at least 2.5 eV.
Correspondingly, $\Delta_{s}W(\omega)$ is a positive constant in
this region (Fig.\ref{Fig4}c), showing no trend to vanish right
above 2.5 eV. Therefore, the scenario where the narrowing of the
Drude peak is fully responsible for the observed SC-induced
increase $\Delta_{s}W(\omega)$, requires an assumption that a
large portion of the Drude peak extends to energies well above 2.5
eV. Given the fact that the bandwidth is about 2 eV, such a
scattering rate seems to be unrealistically large. Accordingly,
$\Delta_{s}W(\omega)$, which corresponds to the discussed
Drude-narrowing model (dotted line in Fig.\ref{Fig5}b) accounts
for only about one-third of the actual value at the cut-off energy
$\Omega_{c}$. It suggests that, at least in optimally doped
Bi2212, a more plausible explanation is a {\it
superconductivity-induced spectral weight transfer from the
interband transitions to the intraband peak}.

While the redistribution of SW below 2.5 eV is experimentally well
determined, the observation of the details of the interband
spectral weight removal, which is most likely spread over a very
broad range of energies, is beyond our experimental accuracy at
the moment.
\subsection{The difference between Bi$_{2}$Sr$_{2}$CaCu$_{2}$O$_{8}$ and YBa$_{2}$Cu$_{3}$O$_{6.9}$}
The results of this article, based on our full data analysis,
refer to Bi2212 near optimal doping. It is interesting to analyze
the picture of spectral weight transfer in other compounds. In
Ref.\onlinecite{BorisScience04} the results of ellipsometric
measurements on detwinned single crystal of optimally doped Y123
were reported. The authors conclude that the intraband spectral
weight decreases in the SC state, following exactly the same
reasoning, as in the case of Bi2212.

Although we think that the model-independent arguments of
Ref.\onlinecite{BorisScience04} are not justified (see Section
\ref{KK}), and actually fail to give the right answer in the case
of Bi2212, we do not rule out the possibility that the spectral
weight transfer in Y123 might be quite different from Bi2212. An
experimental indication of such a possibility is that the
temperature-dependent curves of $\epsilon_{1}(T)$ show an upward
kink at \Tc\ (see Fig.1c-d of Ref.\onlinecite{BorisScience04}),
while in our data of Bi2212 the kink is downward\cite{Text4}. Our
data on twinned Y123 films at optimal doping (\Tc\ = 91 K) show a
similar effect \cite{HajoThesis}. The approximate formula
(\ref{swapprox}) suggests that the sign of
$\Delta_{s}W(\Omega_{c})$ might be different in the two compounds.
However, a more careful analysis is needed for definite
conclusions, especially because of strong temperature-dependent
interband transitions in Y123 around 1 - 2 eV
\cite{HajoThesis,BorisScience04}.

A striking feature of Y123, is that along the direction of the
chains (b-axis) the upward kink of $\epsilon_{1}(\omega)$ at \Tc\
($\Delta_{s}\epsilon_{1}(\omega)$) is much larger than
perpendicular to the chains (compare Figs. 1 D and S3 B of
Ref.\onlinecite{BorisScience04}). It suggests that the charge
dynamics in the chains, or even, the charge redistribution between
the chains and the planes\cite{KhomskiiPRB92} has a strong
influence on the SC-induced spectral weight transfer.
\section{Summary}
We presented a detailed analysis of the optical data, published
earlier in Ref.\onlinecite{MolegraafScience02}. By taking
advantage of a high temperature resolution, we determine the kinks
(slope changes) at \Tc\ of directly measured optical quantities -
reflectivity $R(\omega)$ below 0.75 eV and ellipsometrically
measured $\epsilon_{1}(\omega)$ and $\epsilon_{2}(\omega)$ at
higher energies. The Kramers-Kronig constrained modelling of the
slope-difference spectra clearly shows an extra gain of the
low-frequency integrated spectral weight a result of the
superconducting transition. This gain is not compensated at 2.5 eV
and somewhat higher energies, which suggests, that it is mostly
caused by the spectral weight transfer from the interband towards
intraband transitions and only partially by the narrowing of the
Drude peak.

We found no serious discrepancies between the experimental data of
Refs. \onlinecite{BorisScience04} and
\onlinecite{MolegraafScience02}, insofar they relate to the same
compound (Bi2212). In our opinion, the opposite conclusions drawn
by the authors of Ref.\onlinecite{BorisScience04} are, at least in
part, caused by an incorrect data analysis.

As a concluding remark, the picture of the SC-induced spectral
weight transfer in cuprates is far from being completed. A recent
study \cite{DeutscherCM05} suggests that in the overdoped regime
the spectral weight transfer is conventional (BCS-like), while it
is unconventional (opposite to BCS-like) in the optimally- and
underdoped side. It is also not clear how individual features of
certain compounds (for example, chains in
YBa$_{2}$Cu$_{3}$O$_{6+x}$, structural distortions {\it etc.})
affect this subtle effect. Further experiments should clarify this
issue.

\acknowledgements The authors wish to thank A. F. Santander-Syro,
N. Bontemps, G. Deutscher, J. Orenstein, M. R. Norman, B. Keimer,
C. Bernhard and A. V. Boris for fruitful discussions. This work
was supported by the Swiss National Science Foundation through the
National Center of Competence in Research "Materials with Novel
Electronic Properties-MaNEP".
\section*{Appendix A1. Determination of $\Delta_{s}R(\omega)$ and
$\Delta_{s}\epsilon_{1,2}(\omega)$}
According to the definition (\ref{kinkdef}), the determination of
$\Delta_{s}R(\omega)$, involves the calculation of $\partial
R(T)/\partial T$ above and below \Tc. The numerical derivatives,
calculated straightforwardly from the data points shown in
Fig.\ref{Fig2}, are rather noisy, with the exception of some
frequencies, where the signal is especially good. In order to
limit the statistical noise, we can take advantage of the large
number of temperatures measured, and use the following procedure.
For each frequency, a curve $R(T)$ is fitted to a second-order
polynomial $P_{low}(T)$ between $T_{low}$ and \Tc, and another
polynomial $P_{high}(T)$ between \Tc\ and $T_{high}$, where
$T_{low} < T_{c}$ and $T_{high} > T_{c}$ are some selected
temperatures. For the optimally doped Bi2212 we used $T_{low}$ =
30 K, $T_{c}$ = 88 K and $T_{high}$ = 170 K. The polynomial curves
below and above \Tc\ are shown by respectively blue and red curves
in Fig.(\ref{Fig2}). The superconductivity-induced slope change is
calculated as \begin{equation} \Delta_{s}R =
\left.\frac{dP_{high}(T)}{dT}\right|_{T_{c}} -
\left.\frac{dP_{low}(T)}{dT}\right|_{T_{c}}. \end{equation}
We estimate the error bars of $\Delta_{s}R$ by varying $T_{low}$
and $T_{high}$ in certain reasonable limits (20-40 K and 150 - 200
K respectively). These error bars reflect mostly the systematic
uncertainties of the numerical procedure.
Exactly the same method was applied to determine
$\Delta_{s}\epsilon_{1}(\omega)$ and
$\Delta_{s}\epsilon_{2}(\omega)$ from the temperature-dependent
curves, shown in Fig.\ref{Fig3}.

\begin{figure}[h!]
   \includegraphics[width=8cm,clip=true]{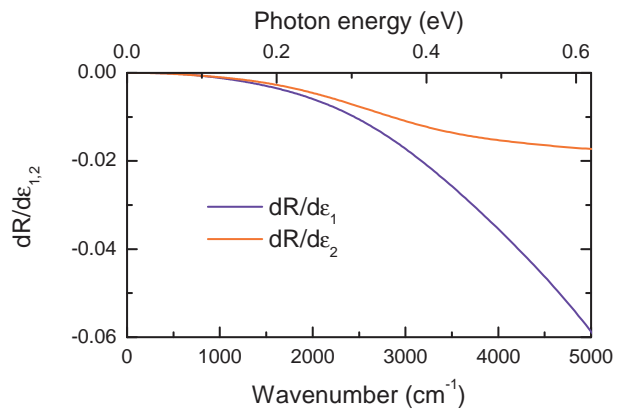}
   \caption{The 'sensitivity functions' $(\partial R/\partial
   \epsilon_{1})(\omega)$, and $(\partial R/\partial
   \epsilon_{2})(\omega)$ in the near- and far-infrared, obtained for Bi2212 near optimal doping
   at \Tc = 88 K as described in Appendix A2.
   }
   \label{Fig6}
\end{figure}

\section*{Appendix A2. Determination of $(\partial R/\partial
\epsilon_{1,2})(\omega,T_{c})$}
The application of the slope-difference analysis in the range
where only $R(\omega)$ is measured requires the knowledge of
'sensitivity functions' $(\partial R/\partial
\epsilon_{1,2})(\omega)$, taken at \Tc. We determine them from a
Drude-Lorentz model, which fits well both $R(\omega)$ and the
ellipsometrically measured $\epsilon_{1}(\omega)$ and
$\epsilon_{2}(\omega)$ at higher frequencies. The accuracy in this
case is superior to the usual KK transform of reflectivity, since
the high-frequency ellipsometry data effectively 'anchor' the
phase of $R$ at low frequencies \cite{BozovicPRB90,KuzmenkoCM05}.
The result is shown in Fig.\ref{Fig6}. Both $(\partial R/\partial
\epsilon_{1})(\omega,T_{c})$ and $(\partial R/\partial
\epsilon_{1})(\omega,T_{c})$ are rather structureless and
vanishing, as $\omega$ goes to 0.


\begin{thebibliography}{100}
\bibitem{signconvention}
The definition of $\Delta\sigma_1(\omega)$ being positive to
indicate a decrease of $\sigma_1(\omega)$ below \Tc\ is somewhat
confusing. In this work we use the opposite definition.
\bibitem{Text1}
The KK relation was already a crucial argument in the analysis,
presented in Ref.\onlinecite{MolegraafScience02}.
\bibitem{footnotekink} In Bi2212, one observes that the
temperature dependent quantities such as the entropy, $S(T)$, do
not have a real kink, but rather an inflection point at $T_c$.
Correspondingly the specific heat, $dS/dT$, has a lambda-like
maximum at $T_c$ instead of a sharp specific heat jump. Fitting
the temperature dependence over some finite temperature range
above and below $T_c$ with a second order polynomial, and taking
the slope difference as we do, measures the slope change over the
fluctuation region around $T_c$. If the fluctuation region becomes
too broad, it becomes increasingly difficult to separate the
superconductivity induced effects from the normal state trends.
For Bi2212 near optimal doping, which is the case which we
consider here, the fluctuation region is sufficiently narrow (see
Ref.\onlinecite{CorsonNature99}), allowing an unproblematic and
meaningful kink-analysis.
\bibitem{Text6}
The only requirement for the KK-consistency of the dispersion
model (\ref{DL}) is that all $\gamma_{i} \geq 0$.
\bibitem{Text4} We note, that the symbol size of the curve
$\epsilon_{1}(T)$, reported in Ref.\onlinecite{BorisScience04} for
Bi2212 (Fig. S6 A) does not exclude a slight downward kink at \Tc,
in agreement with our data.
\end{thebibliography}
\end{document}